\documentclass[useAMS,usenatbib]{mn2e}

\usepackage{graphics,graphicx,url}
\usepackage{amssymb,amsmath}
\usepackage[pass,a4paper]{geometry}
\usepackage{bm}
\usepackage{color}

\newcommand{\beq}{\begin{equation}}
\newcommand{\eeq}{\end{equation}}
\newcommand{\vb}[1]{\textbf{\em #1}}

\title[Glitches in PSR~J1119$-$6127]{The unusual glitch recoveries of the high magnetic field pulsar J1119--6127}
\author[D.~Antonopoulou et al.]{D.~Antonopoulou,$^{1}$\thanks{E-mail: antonopoulou.danai@gmail.com}
P.~Weltevrede,$^{2}$
C. M.~Espinoza,$^{3 }$  
A. L.~Watts,$^{1}$
S.~Johnston,$^{4}$
\newauthor 
R. M.~Shannon$^{4}$ and 
M.~Kerr$^{4}$\\
$^{1}$Astronomical Institute Anton Pannekoek, University of Amsterdam, Postbus 94249, 1090GE Amsterdam, The Netherlands. \\
$^{2}$Jodrell Bank Centre for Astrophysics, School of Physics and Astronomy,
The University of Manchester, Manchester M13 9PL, UK. \\
$^{3}$Instituto de Astrof\'isica, Facultad de F\'isica, Pontificia Universidad Cat\'olica de Chile, Casilla 306, Santiago 22, Chile. \\
$^{4}$CSIRO Astronomy and Space Science, Australia Telescope National Facility, PO Box 76, Epping, NSW 1710, Australia.\\
}

\date{\today}
\pagerange{\pageref{firstpage}--\pageref{lastpage}} \pubyear{2012}
\begin{document}

\maketitle
\label{firstpage}

\begin{abstract}

Providing a link between magnetars and radio pulsars, high magnetic field neutron stars are ideal targets to investigate how bursting/magnetospheric activity and braking torque variations are connected to rotational glitches. 
The last spin-up glitch of the highly magnetised pulsar J1119$-$6127 back in 2007 was the first glitch in a rotationally powered radio pulsar to be accompanied by radiative changes. Moreover, it was followed by an uncommon glitch relaxation that resulted in a smaller spin-down rate relative to the prediction of the pre-glitch timing model. 
Here, we present 4 years of new radio timing observations and analyse the total of 16 years of timing data for this source. 
The new data uncover an ongoing evolution of the spin-down rate, thereby allowing us to exclude permanent changes in the external or internal torque as a stand-alone cause of the peculiar features of the glitch recovery. 
Furthermore, no additional variations of the radio pulse profile are detected, strengthening the association of the previously observed transient emission features with the glitching activity. 
A self-consistent measurement of the braking index yields a value $n\simeq2.7$, indicating a trajectory in the $P-\dot{P}$ plane inclined towards the magnetars.
Such a potential evolutionary link might be strengthened by a, possibly permanent, reduction of $\sim15\%$ in $n$ at the epoch of the 2007 glitch.

\end{abstract}


\section{Introduction}

Studying the rotational dynamics of radio pulsars is of key importance for the advance of our global understanding of neutron stars (NSs). 
Isolated NSs slowly brake as they lose energy through electromagnetic torques in their magnetospheres. 
In the simplest approximation, the spin-down is described by the dipole radiation of a misaligned rotator in vacuum, $\dot{\nu}\propto\nu^3$, where $\nu$ is the NS spin frequency and dots represent time derivatives.
However, the less understood contribution from magnetospheric currents can be significant or even dominant. 
It is conventionally presumed that a generalised spin-down law 
\beq
\label{obs:powerlaw}
\dot{\nu}=-C\nu^n
\eeq
governs the rotation which, under the assumptions that $n$ and the positive factor $C$ are time-independent, leads to the observable 
\beq
\label{eq:brakingindex}
n=\nu\ddot{\nu}/\dot{\nu}^2 \,\text{,}
\eeq
the so-called braking index. 
According to the prediction of the dipole in vacuum braking mechanism, $n=3$ if $C$ can be considered constant for most of the pulsar's life. 
In this case $C$ depends on the magnetic dipole moment and the moment of inertia of the stellar component that follows the spin-down. 
Measurements of $n$ can probe the contribution of other braking mechanisms in the spin-down rate, as well as the validity of the generalised spin-down law and the underlying assumptions for $C$ and $n$. 
There are only a few pulsars for which a reliable measurement of $\ddot{\nu}$ is possible, and in all of them deviations from the rudimentary vacuum dipole braking are observed, with $n<3$ (see for example Table 1 in \citet{ljg+14} and references therein).

Another generic feature of NS rotation is timing noise. 
It manifests as small fluctuations of the star's frequency with respect to a simple spin-down model, on time scales of days to years. 
The nature of timing noise remains unknown, though recent work relates it to magnetospheric phenomena such as pulse profile changes \citep{lhk+10}. 
Finally, occasional abrupt spin-ups called glitches have been observed in $\sim 160$ pulsars. 
The increase in spin frequency $\nu$ during a glitch happens very rapidly and typically lies in the range  $(10^{-5}\,-100)\;\rm{\mu}\rm{Hz}$. 
Glitches are usually accompanied by an increase in spin-down rate $|\dot{\nu}|$ of the order of $(10^{-19}-10^{-11})\;\rm{Hz}\,\rm{s^{-1}}$, often followed by a slow relaxation towards the pre-glitch rotational parameters  \citep{elsk11,Yu13}.

Large glitches in rotationally powered pulsars (RPPs) have not been connected to observed radiative changes in either intensity, polarisation or pulse profile shape, which supports an internal, rather than magnetospheric, origin. 
Our understanding of pulsar glitches is far from complete, however a two-component model for the NS's interior has been successfully implemented to describe their main attributes  \citep{ai75,aaps84a,hps12}. 
In such models, a neutron superfluid component, usually assumed to be the inner crust's superfluid, is at least partially decoupled from the rest of the star and rotates faster than the normal component. 
The latter encompasses the solid outer and inner crust and all fluid and superfluid components that are strongly coupled to it, and is spinning down under the external torques. 
When angular momentum is rapidly exchanged between the two components, a glitch occurs. 
The spin-up brings the two components closer to corotation, which weakens their coupling. 
As a result the external torque acts on a reduced ef{\kern0pt}fective moment of inertia $I_{\rm{ef{\kern0pt}f}}$, leading to an enhanced spin-down rate following the glitch.
The recoupling process of the superfluid is reflected in the post-glitch relaxation, which can often be described as exponential with long characteristic timescales, from days to months \citep{sl96}.

Though most common in young radio pulsars, glitches appear in other RPPs like millisecond pulsars and old, slower pulsars too \citep{cb04,elsk11}. 
Glitches have also been observed in the rotation of magnetars, NSs which present X-ray luminosities that exceed their rotational energy losses and bursting activity such as very energetic $\gamma$-ray flares or X-ray outbursts. 
Magnetars are thought to be highly magnetised NSs, powered by the decay of their strong magnetic fields. 
Contrary to ordinary radio pulsars, glitches in magnetars often (but not always) coincide with bursts or smaller radiative changes \citep{dibkaspi14}. 
Furthermore, magnetars show a larger variety of post-glitch recoveries as well as other spin-down rate fluctuations, sometimes accompanying a radiative event but without an apparent glitch association \citep[see for example][]{wkv+99,gdk09,antiglitch13,dibkaspi14}. 
PSR J1846$-$0258, which is normally powered by rotation, exhibited magnetar-like activity \citep{ggg+08} together with a large glitch which was followed by an atypical $\dot{\nu}$ evolution, indicative of a possibly permanent decrease in the braking index \citep{kh09,lkg10,lnk+11}.

PSR~J1119$-$6127 is a high magnetic field pulsar, a small class of RPPs with spin parameters and inferred magnetic field strengths close to those of magnetars. 
It was first discovered in the radio during the Parkes Multibeam Pulsar Survey and has a period of $P=408\,\rm{ms}$ and period derivative $\dot{P}=4\times10^{-12}$ \citep{ckl+00}. 
Its surface dipole magnetic field $B_d$ strength is estimated, assuming conventional dipole braking, as $B_d\simeq4.1\times10^{13}\,\rm{G}$, very close to the quantum electrodynamics (QED) limit $B_{QED}=4.41\times10^{13}\,\rm{G}$ above which phenomena like spontaneous pair creation and suppression of pair cascades due to photon splitting must be taken into account \citep{bh01}. 

The characteristic age $\tau_{sd}=P/2\dot{P}\sim1.6-1.9\,\rm{kyr}$ of PSR~J1119$-$6127 suggests it is a young pulsar, and indeed it has been associated with the supernova remnant SNR G292.2-0.5 \citep{cgk+01}. 
It is one of the youngest radio pulsars with detected thermal emission, as it was discovered  in X-ray observations by \citet{pkc+01} and since then observed by several missions like {\it ASCA}, {\it ROSAT},{\it XMM} and {\it Chandra} \citep{gs03b,gkc+05,ng12}.
Pulsations were also recently detected by {\it Fermi}, as expected from its high rotational energy loss rate ($\dot{E}=2.3\times10^{36}\,\rm{erg}\,\rm{s}^{-1}$), making PSR~J1119$-$6127 the source with the highest inferred $B_d$ among $\gamma$-ray pulsars \citep{pnt11}. 
The relatively stable spin-down and long-term monitoring with the Parkes radio telescope allowed the accurate measurement of its braking index, which was found to be $n=2.684\pm 0.002$ \citep[hereafter WJE11]{wje11}.
  
In 2004, PSR~J1119$-$6127 suf{\kern0pt}fered a rather common glitch of magnitude $\Delta\nu_g\simeq0.7\,\rm{\mu Hz}$ (where an index $g$ indicates values extrapolated at the glitch epoch), accompanied by a change in spin-down of $\Delta\dot{\nu}_g\simeq-9\times10^{-14}\,\rm{Hz\,s^{-1}}$ which recovered with a characteristic timescale of $\sim3$ months. 
However the glitch recovery was unusual, in that the post-glitch spin-down rate appeared to settle at a smaller value than the one extrapolated from the pre-glitch spin parameters.
This is very atypical of glitches, where either the pre-glitch spin-down rate resumes once the post-glitch relaxation is over, or the end result is a larger spin-down rate which does not appear to recover completely.
The evolution of this anomalous $\dot{\nu}$ recovery was interrupted in 2007 by another glitch. 
The second glitch had a much larger magnitude, with $\Delta\nu_{g}>10\,\rm{\mu Hz}$ and $-10^{-10}\lesssim\Delta\dot{\nu}_{g}\lesssim-10^{-11} \,\rm{Hz/s}$ inferred at the glitch epoch, showing a relaxation on two timescales ($\sim10$ days and $\sim6$ months). 
As in the first glitch, the post-glitch spin-down rate eventually decreased below the value extrapolated from the pre-glitch timing model, showing a maximum departure of $\Delta\dot{\nu}_{max}\simeq3.5\times10^{-14} \,\rm{Hz/s}$. 
Even more surprisingly, the first radio observation after this glitch showed a very dif{\kern0pt}ferent pulse profile and erratic pulse components. 
The magnetospheric activity was present only after the glitch and lasted no more than $\sim3$ months, a strong indication that the two phenomena might be related (WJE11).
This remarkable behaviour offers a unique opportunity to explore the connection of internal and external processes during glitches.

In this paper we present the latest timing observations of PSR~J1119$-$6127 and a self-consistent method to analyse its braking index. The new data show that the magnitude of the spin-down rate has been relatively {\it increasing} for the last $\sim5$ years, meaning that the difference with the pre-glitch predicted $\dot{\nu}$ is reducing.  
We discuss the follow up of the puzzling $\dot{\nu}$ evolution after the 2007 glitch in view of current theories of glitches and pulsar spin-down.


\section{Observations and timing analysis}
\label{observations}

For this work we analyse 16 years of timing observations of PSR~J1119$-$6127.
This is the data analysed in WJE11, supplemented
with four years of new data obtained from the ongoing timing
observations with the 64-m Parkes radio telescope in Australia.
In this timing program \citep{wjm+10} each pulsar is typically
observed once per month at a wavelength of 20~cm and twice per year at
10 and 50~cm. The timing analysis detailed in WJE11 was
repeated for this longer data set. This process is summarised below
while for details we refer the reader to the aforementioned papers.

The individual observations were summed resulting in a high
signal-to-noise ``standard'' profile (see also \citealt{wj08b}).  The
time-of-arrival (TOA) of each observation was determined by
cross-correlation of their pulse profile with this standard. These
TOAs were projected to the solar system barycenter using the TEMPO2
timing package \citep{hem06}. Timing analysis of these corrected TOAs was performed using custom
software (see WJE11).

The basic timing model used to describe the
effect of spin-down on the rotational phase of the pulsar as a function
of time $\phi(t)$ is a truncated\footnote{However, see section \ref{brakingindex} for the slightly different, but self-consistent, way to model the phase in terms of the generalised power law spin-down.} Taylor series:
\begin{align}
\label{EqBasicTimingModel}
\phi(t)=&\phi_0+\nu_0\cdot(t-t_0)+\frac{\dot{\nu}_0}{2}\cdot(t-t_0)^2+\frac{\ddot{\nu}_0}{6}\cdot(t-t_0)^3.
\end{align}
Here $\phi_0$, $\nu_0$, $\dot{\nu}_0$ and $\ddot{\nu}_0$ are the reference phase, spin frequency and its first two time derivatives defined at epoch $t_0$.
Each glitch was modelled by including 
an additional function $\phi_\mathrm{g}(t)$ to the timing model after the glitch epoch $t_g$, which is parameterised as
\begin{align}
\nonumber \phi_\mathrm{g}(t)=&\Delta\phi+\Delta\nu_p\cdot(t-t_g)+
                    \frac{\Delta\dot{\nu}_p}{2}\cdot(t-t_g)^2+\frac{\Delta\ddot{\nu}_p}{6}\cdot(t-t_g)^3\\
\label{EqTimingModelGlitch}
                     &-\left(\sum_i\Delta\nu_d^{(i)}\tau_d^{(i)} e^{-(t-t_g)/\tau_d^{(i)}}\right),
\end{align}
where $\Delta\phi$ is a phase offset arising from the fact that the glitch epoch is not accurately known.  
The parameters labelled with an index $p$ correspond to permanent changes while those with an index $d$ refer to decaying components of the spin evolution.
The full set of parameters in Eq. \ref{EqTimingModelGlitch} is degenerate for our dataset, so each glitch was modelled with a subset of these terms as described in the next subsection. The $\Delta\ddot{\nu}_p$ term was not required in the description of the shorter data-span presented in WJE11 and was set to zero in their timing model.

\subsection{Spin evolution}
\label{sect:spinevolution}

\begin{figure*}
\begin{center}
\includegraphics[height=0.99\hsize,angle=270]{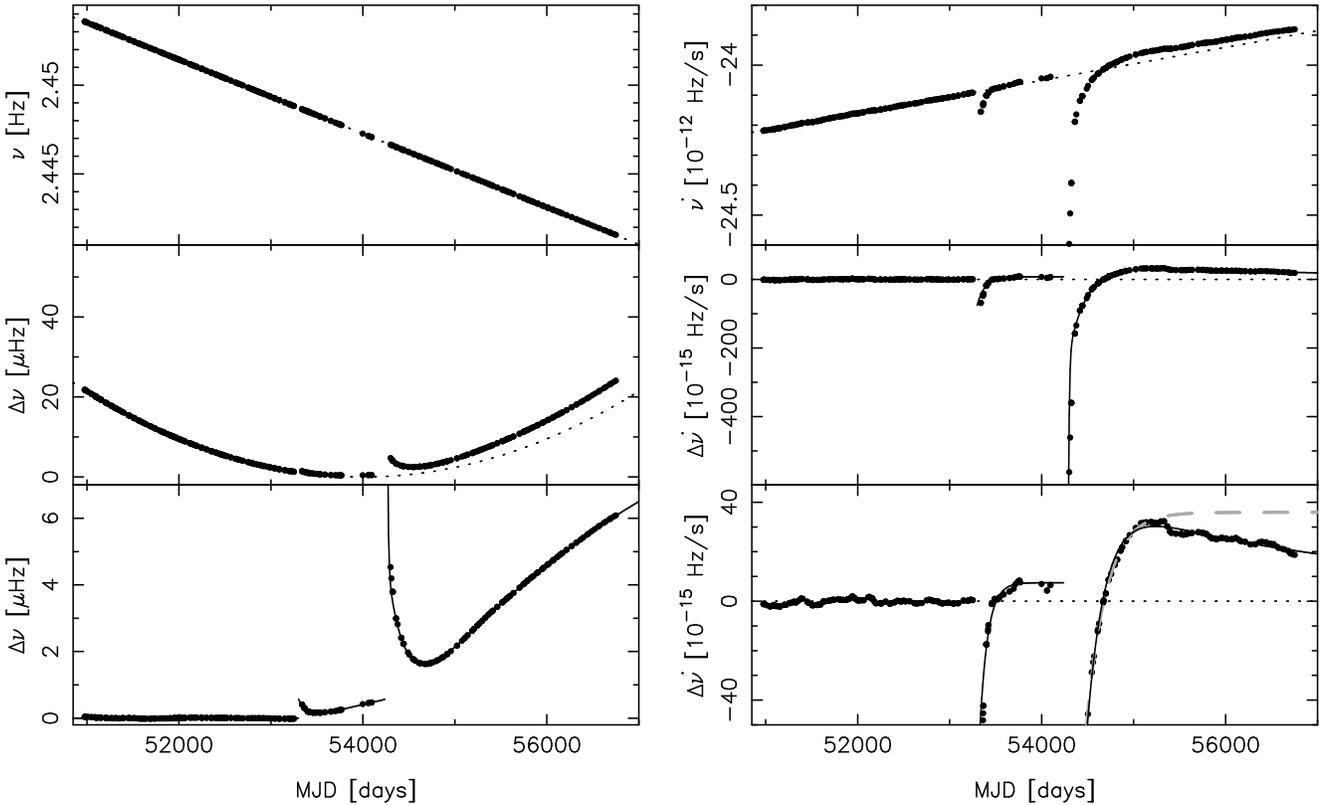}
\caption{\label{FigSpinEvolution}{\em Top-left panel:} The measured
  rotational frequency of PSR~J1119$-$6127 as a function of time (error
  bars are smaller than the points). The frequency is steadily
  decreasing, as is expected from the measured $\nu_0$ and $\dot{\nu}_0$
  in Table \ref{TableTimingSolution} (dotted line). {\em Middle-left
    panel:} The effect of a constant spin-down rate is subtracted from
  the rotational frequency. Especially the second glitch can clearly
  be seen as a deviation from the parabola-shape. The latter indicates
  a significant and stable braking-index and the prediction according
  to the measured $\ddot{\nu}_0$ is shown as a dotted line. {\em
    Bottom-left panel:} The difference between the measured spin
  frequency and the contributions from $\nu_0$, $\dot{\nu}_0$ and
  $\ddot{\nu}_0$. The solid lines show the prediction according to the
  glitch model A. {\em Top-right panel:} The measured spin-frequency
  derivative as a function of time compared to the $\dot{\nu}_0$ and
  $\ddot{\nu}_0$ contribution (dotted line). {\em Middle-right panel:}
  The difference between the data and the dotted line of the top-right
  panel. The solid line indicates the prediction according to the
  glitch model A.  {\em Bottom-right panel:}
  This plot is identical to the middle-right panel, but using a
  slightly more constrained vertical range. The timing model from
  WJE11 is indicated as the grey dashed line. Although these plots show theoretical curves according to predictions of model A, both models in Table \ref{TableTimingSolution} are
  virtually identical over the time span covered by
  observations. }
\end{center}
\end{figure*}

The rotational evolution (the spin frequency and the spin-down rate) of
PSR~J1119$-$6127 is shown in Fig. \ref{FigSpinEvolution}, which is
essentially an update of Fig. 10 in WJE11. These parameters
were measured for illustrative purposes by fitting a timing model to short stretches of data.
For each TOA we defined a stretch of data that included all TOAs separated by less than 75 days
from the central TOA. Stretches of data containing less than four TOAs
were excluded from the analysis.

The rotation of PSR~J1119$-$6127 is slowing down over time with an approximately constant rate, parameterised by $\dot{\nu}_0$. 
This results in the steep gradient in the spin frequency against time as observed
in the top left panel of Fig. \ref{FigSpinEvolution}. 
However, as a result of the significant and stable $\ddot{\nu}$ (WJE11), the shape is
better described by a parabola (parameterised by $\ddot{\nu}_0$). 
This is revealed after subtracting the effect of $\dot{\nu}_0$ (middle-left
panel). 
The main deviations from an otherwise almost perfect parabola are the two glitches and their recoveries. 
These deviations are more pronounced after
subtracting the effect of the long-term spin evolution, parametrised by $\dot{\nu}_0$ and $\ddot{\nu}_0$ (bottom-left panel). 

The two unambiguous glitches at MJD $\sim$53290 and $\sim$54240 are characterised by a
sudden spin-up, followed by a gradual
recovery which can be parameterised in terms of a model for
the evolution in $\dot\nu$ (top-right panel of
Fig. \ref{FigSpinEvolution}).
This evolution is dominated by a linear increase as a result of
the stable $\ddot{\nu}_0$ (dotted line). 
The deviations ($\Delta\dot\nu$) caused by the two glitches and their recoveries 
are presented (zoomed-in) in the middle and bottom right panels.
Here $\Delta\dot\nu$ is the difference of $\dot\nu$ from what we will refer to as the ``projected spin-down
rate'', corresponding to the dotted line in the top-right panel
(i.e. the expected $\dot\nu$ evolution for a constant $\ddot\nu$).  

Both glitches presented a sudden increase in spin-down rate
followed by an exponential recovery, which is a common feature of
glitching behaviour.  Based on data up to MJD 55364, WJE11
reported that the recoveries were such that $\dot{\nu}$ overshoots
the projected pre-glitch spin-down rate, resulting eventually in a
{\em slower} spin-down rate.  Thus the initially negative 
$\Delta\dot\nu$ evolves to a positive, persisting
$\Delta\dot\nu>0$, about $\sim0.5$ and $\sim1.5$ years after the
1\textsuperscript{st} and 2\textsuperscript{nd} glitch respectively
(see bottom-right panel).  As discussed in WJE11, this
evolution is not normal among the rest of the glitch population.

The new data (after MJD 55364) reveal that the described picture is
incomplete for the second, larger glitch. The end of the data-span
used in WJE11 happens to correspond roughly to the date at
which $\Delta\dot\nu$ peaks in the bottom-right panel of
Fig. \ref{FigSpinEvolution}. Clearly the post-glitch recovery is not
converging to a permanent and constant positive value of
$\Delta\dot\nu$.
Instead, the currently positive $\Delta\dot\nu$ is slowly decreasing, and the spin-down rate evolves towards the projected pre-glitch $\dot\nu$.
This newly discovered long-term evolution cannot be modelled with the
set of parameters of the timing solution presented by WJE11,
hence an additional term is required. 
The functional form of this term
is not a priori known. Here we explore two different parameterisations
to model the long-term recovery of the second glitch.

\begin{table}
\caption{Rotational parameters for PSR~J1119$-$6127 according to two different models describing the 2007 glitch and its recovery. Model A includes three exponential recovery terms and a permanent change only in $\nu$, while in model B the longest timescale exponential recovery is replaced with a permanent change in $\dot\nu$ and $\ddot\nu$. Although all model parameters were optimised simultaneously for the whole dataset, only the parameters describing the 2007 glitch are significantly different for these two models.}
\label{TableTimingSolution}
\begin{center}
\begin{tabular}{lcc}
\hline 
Parameter  \hspace{1.7cm}                 &       Model A   & Model B\\
\hline
Epoch (MJD)                               &       54000           &  54000  \\
$\nu_0$~(Hz)                                &   2.447266543(6)      &  2.447266540(6)          \\
$\dot{\nu}_0$~($10^{-15}~$Hz\,s$^{-1}$)      &   $-$24050.94(7)        &  $-$24050.97(8)              \\
$\ddot{\nu}_0$~($10^{-24}~$Hz\,s$^{-2}$)     &   637.4(4)            &  637.2(5)                 \\
DM~(cm$^{-3}$\,pc)                        &       713             & 713 \\
$n$                                        &   2.677(2)$^\dagger$ & 2.677(2)$^\dagger$\\
MJD range                                 & 50850\,--\,56794  &       50850\,--\,56794 \\
RMS residuals (ms)                        &   68.6            &    74.8\\
        &               \\[-2mm]
\multicolumn{3}{c}{\bf 2004 glitch parameters} \\
        &               \\[-2mm]
Glitch epoch                                     &   53290    &    53290   \\
$\Delta\nu_p$~($\mu$\,Hz)                        &  $-$0.05(2) &    $-$0.04(2) \\
$\Delta\dot{\nu}_p$~($10^{-15}~$Hz\,s$^{-1}$)     &   7.5(4)    &    7.3(4)   \\
$\Delta\nu_d$~($\mu$\,Hz)                        &   0.70(5)   &    0.71(6)  \\
$\tau_d$~(days)                                  &   86(11)    &    80(11)   \\
        &               \\[-2mm]
\multicolumn{3}{c}{\bf 2007 glitch parameters} \\
        &               \\[-2mm]
Glitch epoch                                   &   54240   &  54240   \\
$\Delta\nu_p$~($\mu$\,Hz)                      &   6.5(2)  & $-$0.91(2)    \\
$\Delta\dot{\nu}_p$~($10^{-15}~$Hz\,s$^{-1}$)   &   --      & 32.4(4)     \\
$\Delta\ddot{\nu}_p$~($10^{-24}~$Hz\,s$^{-2}$)  &       --  & $-$93(2)   \\
$\Delta\nu_d^{(1)}$~($\mu$\,Hz)                &    5.80(4) & 5.66(5)     \\
$\tau_d^{(1)}$~(days)                          &    194(2)  & 184(2)      \\
$\Delta\nu_d^{(2)}$~($\mu$\,Hz)                &    75(200) & 80(200)       \\
$\tau_d^{(2)}$~(days)                          &    10(5)   & 9(7)      \\
$\Delta\nu_d^{(3)}$~($\mu$\,Hz)                &    $-$7.7(2) & --         \\
$\tau_d^{(3)}$~(days)                          &    2324(60) & --        \\
\hline
\end{tabular}
\end{center}

$^\dagger$ The quoted measurements for the braking index do not follow directly from the spin parameters of the two timing solutions, but from a self-consistent analysis of pre-glitch data only, as explained in section \ref{brakingindex}.

\end{table}

In the timing solution presented by WJE11, the 2007 glitch recovery is parameterised by instantaneous changes in both $\nu$ and $\dot\nu$: the permanent $\Delta\nu_p$ and $\Delta\dot\nu_p$ steps (but without a $\Delta\ddot{\nu}_p$ term) and two exponentially decaying terms which describe the relaxation.
In a way, the most natural extension of this model would be to add a third
exponential recovery term with an amplitude $\Delta\nu_d^{(3)}$ and a
long associated timescale $\tau_d^{(3)}$. 
We refer to this as model A (see Table \ref{TableTimingSolution}). 
There is no need for a permanent jump in $\dot\nu$ in this model, because it can be absorbed in the other fit parameters. 
Hence model A has effectively one extra free parameter compared to the timing solution presented by WJE11. 
Note that $\Delta\nu_d^{(3)}<0$, opposite to the other two decaying terms.
Since the associated timescale $\tau_d^{(3)}$ is very large (about 6
years according to model A), it is currently impossible to distinguish
between an exponential or linear long-term recovery in $\dot\nu$.  
The latter possibility is further explored in what we refer to as model B. 
We model a linear evolution by replacing the third exponential term which was included in model A with the $\Delta\dot{\nu}_p$ and $\Delta\ddot{\nu}_p$ terms. 
Therefore both models have the same number of free parameters.

\begin{figure}
\begin{center}
\includegraphics[height=0.99\hsize,angle=270]{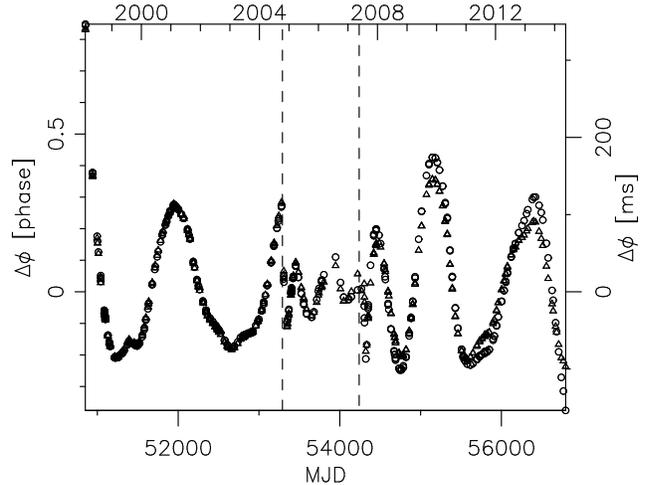}
\caption{\label{FigResiduals}The timing residuals of PSR~J1119$-$6127
  with respect to the timing model A (open triangles) and B (open
  circles) of Table \ref{TableTimingSolution}. The error bars are much
  smaller than the points. The glitch epochs are indicated by the dashed lines.}
\end{center}
\end{figure}

The set of rotational parameters defined in Eqs. \ref{EqBasicTimingModel} and \ref{EqTimingModelGlitch}
were determined simultaneously for the entire dataset by applying
the timing model, including both glitches, directly to the TOAs. 
The values are optimised by minimising
the RMS (root-mean-square) deviation from zero of the timing residuals (the difference of the measured TOA from the one expected from the timing
model, as shown in
Fig. \ref{FigResiduals}). 
For details of the fitting procedure, see WJE11. 

The results are presented in Table \ref{TableTimingSolution}. 
As reflected on the errors of the fast decaying term $\Delta\nu_d^{(2)}$, the initial part of the 2007 glitch relaxation is poorly constrained. This is due to the observing cadence. 
For either model however, the instantaneous spin-up at the glitch epoch appears to be very large, 
$\Delta{\nu}_g\sim80\,\rm{\mu\,Hz}$ ($\Delta{\nu}_g/\nu\sim3\times10^{-5}$), 
placing this glitch among the largest ones ever observed.
The inferred spin-down change at the glitch epoch is also very large, 
comparable to those seen in magnetars, and might be as high as $\Delta\dot{\nu}_g/\dot{\nu}\sim4$.  

\begin{figure}
\begin{center}
\includegraphics[height=0.99\hsize,angle=270]{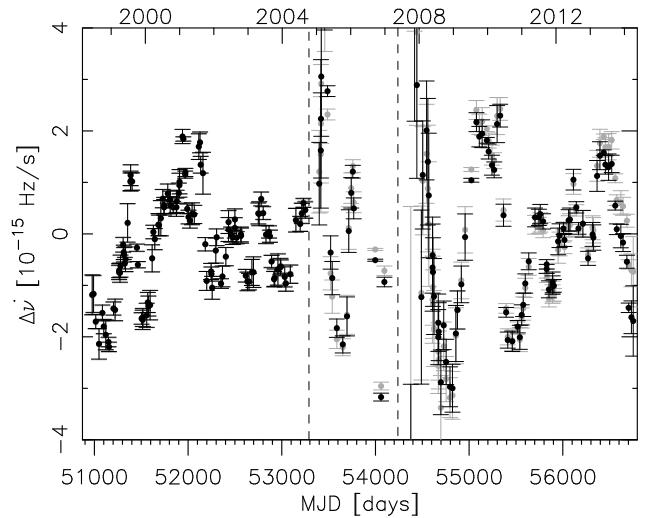}
\caption{\label{FigNuResiduals}The difference between the measured
  spin-down rate of PSR~J1119$-$6127 and the timing models of Table \ref{TableTimingSolution}. 
Black points are used for model A and  grey points for model B. Some points directly after the
  two glitches fall below the plotted range. The glitch epochs are indicated by the dashed lines.}
\end{center}
\end{figure}

There are clear structures in the timing residuals (Fig. \ref{FigResiduals}), indicating that there is
significant timing noise present which is not included in our timing model. 
Although the RMS of the timing residuals is slightly better for model A (see also Table
\ref{TableTimingSolution}), this is not significant within the systematic uncertainties resulting from the timing noise. Observationally it is therefore currently impossible to distinguish between the two models. This is illustrated in Fig. \ref{FigNuResiduals}, which shows the difference between the observed and predicted values of $\dot\nu$ for the two models.

Nevertheless, the extrapolated spin evolution for the two models is quite different. 
By extrapolating model B, it can be predicted that the projected
$\dot\nu$ ($\Delta\dot\nu=0$) according to the solution prior to the 2004 glitch, will be reached again
at MJD $\sim$59178; 13.5 years after the occurrence of the second
glitch. On the other hand, according to model A this value is never reached and $\Delta\dot\nu$
tends to $7.5\times10^{-15} {\rm Hz\,s^{-1}}$, in accordance with the prediction of the post-2004 glitch model. 
We note that the permanent change in $\ddot{\nu}$ included in model B implies a permanent change in the braking index, a possibility which is explored in more detail in section \ref{brakingindex}. 

Only two possible functional forms of the glitch recovery are
explored here. 
In reality the recovery might be quite different and it
could be, for example, a damped oscillation.  
As can be seen in Fig. \ref{FigNuResiduals}, the unmodeled features in the post-glitch $\dot{\nu}$ evolution do not appear to be strictly periodic, similarly to what is observed in the timing residuals (Fig. \ref{FigResiduals}).
However, a simple Fourier transform indicates some excess power at a period of $\sim400$ days, which can perhaps be confirmed with future observations. Notably, \citet{yuan10osc} reported a significant post-glitch oscillation in the residuals of PSR~J2337$+$6151 with a similar period of $364$ days. 
  
It remains to be seen if the long-term glitch recovery for PSR~J1119$-$6127 can be determined
experimentally, as this evolution might be disrupted by another glitch.
It is also not clear if the 2004 glitch presents a similar long-term evolution: the data are too sparse after MJD $\sim$54000 and up to the second glitch epoch to determine whether $\dot\nu$
evolves to a constant $\Delta\dot\nu>0$ or back towards $\Delta\dot\nu=0$.

\subsection{The braking index}
\label{brakingindex}

The long-term $\ddot{\nu}$ is typically unmeasurably small for old pulsars while for young ones it is difficult to determine reliably because the spin-down evolution is dominated by timing noise and glitch recoveries. 
PSR~J1119$-$6127 is one of the only eight pulsars to date for which a long-term, stable $\ddot{\nu}$ and a braking index have been determined \citep{ljg+14}.
In this section we want to measure the long-term $n$, and explore its possible change as a consequence of the 2007 glitch.
Previously, the braking index has been found to be $n=2.91\pm0.05$ by \citet{ckl+00}, and $n=2.684\pm0.002$ by WJE11 who analysed a longer dataset. 
Both these measurements were derived by Taylor expanding the long-term spin-down around a pre-glitch epoch. 
As in Eq. \ref{EqBasicTimingModel}, the free parameters were $\phi_0$, $\nu_0$, $\dot{\nu}_0$ and $\ddot{\nu}_0$ while higher order time derivatives were set to zero. 
The braking index was obtained from the resulting fit parameters by applying Eq. \ref{eq:brakingindex}.
For the dataset analysed by WJE11 additional terms to account for the glitch effects were included in the timing model, as described in section \ref{sect:spinevolution}. 
 
\begin{figure*}
\begin{center}
\includegraphics[height=0.99\hsize,angle=270]{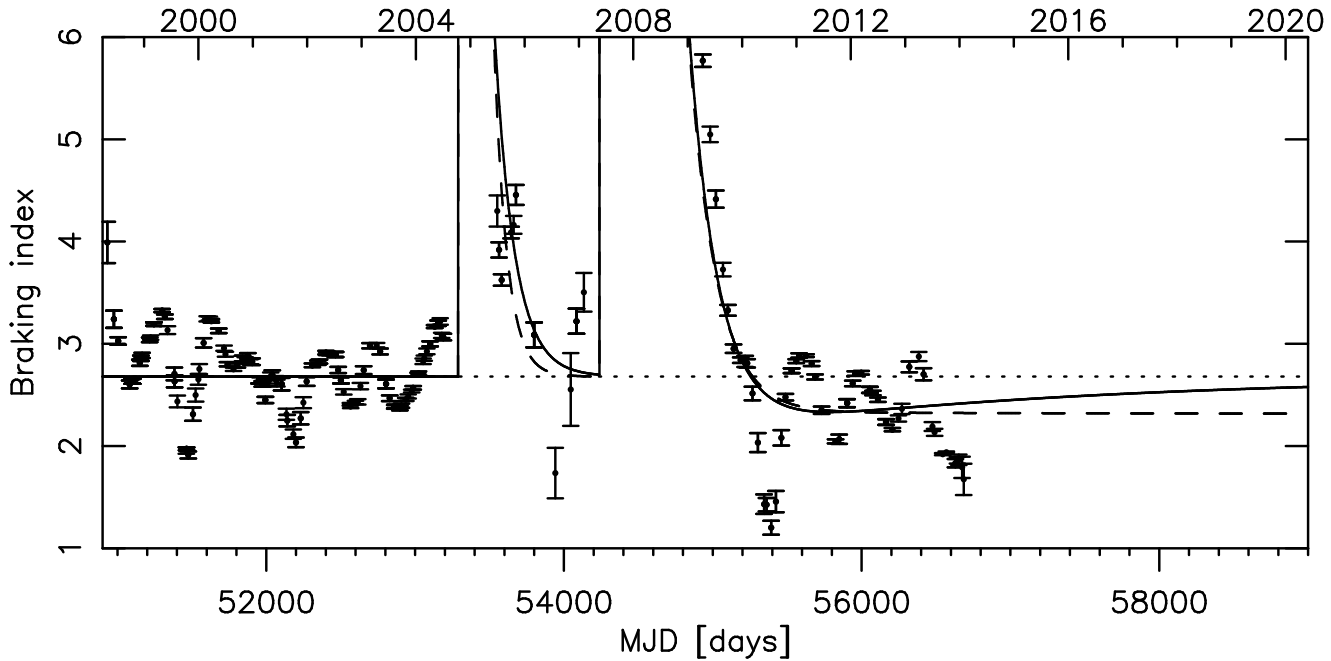}
\caption{\label{FigBreakingIndex} The braking index $n(t)$ of PSR~J1119$-$6127
  obtained from a timing model with a long term power-law spin-down of constant $n$ (short-dashed line) and the two glitches parametrised as in model A (continuous line) and B (long-dashed line). The very high values of $n$ immediately after the glitches are excluded from the plot for clarity. The points are measurements of $n(t)$ from  fits of Eq. \ref{EqBasicTimingModel} to small subsets of data as described in the text. Note that these measurements are highly contaminated by timing noise, not reflected in their statistical error bars plotted here which are much smaller than the overall variations.}
\end{center}
\end{figure*}

However, a fit of the spin-down as in Eq. \ref{EqBasicTimingModel} to a long dataset is not strictly appropriate for the purpose of measuring the long-term $n$. 
This is because Eq. \ref{obs:powerlaw} implies non-zero higher-order frequency derivatives, which are assumed to be negligible in the truncated Taylor approximation (Eq. \ref{EqBasicTimingModel}). 
This affects the measurement of the braking index if the timespan covered by the observations is large enough to invalidate the approximation $\dddot{\nu}_0(t-t_0)\ll\ddot\nu_0$. 
Consequently, when the third and higher order time derivatives (which depend on $t_0$) are ignored, the resulting value of the braking index depends on the arbitrary choice of the reference epoch $t_0$. 
To demonstrate this effect, $t_0$ was varied between MJD 50850 and 53290 (i.e. the time span covered by the data before the 2004 glitch) while fitting for $\nu_0$, $\dot{\nu}_0$ and $\ddot{\nu}_0$ in this range. The derived braking index varied between $n=2.66$ and $2.69$, a variation far greater than the very small error-bar derived via standard error propagation. When $t_0$ is set at the end of the total data-span, $n=2.72$.

To overcome this disadvantage we employed a different, self-consistent method to estimate the underlying braking index.
The measurement of a braking index as in Eq. \ref{eq:brakingindex} is meaningful if the inter-glitch data are indeed described reasonably well by a power law with constant $n$ and $C$. 
Under these assumptions, Eq. \ref{obs:powerlaw} can be integrated twice (from $t_0$ to $t$) and results, for $n\neq\{1\, ,\,2\}$ and using using $C=-\dot{\nu}_0/\nu_0^n$, to the following description of the rotational phase of the star:
\begin{align}
\label{PowerLawTimingModel}
\phi(t)=\frac{\nu_0^2}{\dot{\nu}_0(2-n)}\left\{\left[1+\frac{\dot{\nu}_0}{\nu_0}(1-n)(t-t_0)\right] ^{\frac{2-n}{1-n}}-1\right\}+\phi_0 \;\text{.}
\end{align}
This equation is fully consistent with a spin-down that can be described by the generalised power-law (Eq. \ref{obs:powerlaw}). 
Since in that case Eq. \ref{PowerLawTimingModel} is more precise than a truncated Taylor series, it should provide a more sensitive and robust measurement of the braking index. 
We therefore replaced the long-term timing model (Eq. \ref{EqBasicTimingModel}) by Eq. \ref{PowerLawTimingModel} in our custom timing analysis software,
and performed a fit of the data before the 2004 glitch. 
These fits use the braking index $n$ as a free parameter instead of the long-term $\ddot{\nu}_0$, but have otherwise the same number of fitted parameters and result in residuals similar to those presented in Figure \ref{FigSpinEvolution}. 
The resulting braking index is $n=2.677\pm0.002$, which is the value quoted in Table \ref{TableTimingSolution}. 
This result is independent of the choice of $t_0$. Moreover it is fully consistent with values obtained from the full data-span,
derived by including the glitch model A or B to Eq. \ref{PowerLawTimingModel}. 
 
Measurements of the higher-order frequency derivatives can provide insights to the braking mechanism. 
The generalised power-law (Eq. \ref{obs:powerlaw}) for constant $C$ and $n$
 implies a non-zero $\dddot{\nu}$ for all positive $n\neq0.5$, which should be 
\beq
\label{obs:F3}
\dddot{\nu}=n(2n-1)\dot{\nu}^3\nu^{-2} \;\text{.}
\eeq
Inserting the measured braking index into this equation and the spin parameters of Table \ref{TableTimingSolution} leads to the prediction\footnote{This calculation is dependent on the choice of the reference epoch $t_0$, but not at a level relevant for the arguments made here.} that $\dddot{\nu}_0\sim-2.7\times10^{-32}$ Hz/s$^3$.
By including the higher order $\dddot{\nu}_0$ term in Eq. \ref{EqBasicTimingModel}, and fitting for the pre-glitch data only, leads to $\dddot{\nu}_0=(-4.5\pm0.1)\times10^{-31} {\rm Hz/s^3}$, a measurement which is most likely highly contaminated by unmodelled effects such as timing noise, which are not taken fully into account in the quoted statistical error bars. Since this measurement is an order of magnitude larger than the prediction, it does not provide independent support to the generalised power-law braking hypothesis.

Although timing noise hinders the determination of the
braking index, it should be stressed that the top right panel of
Fig. \ref{FigSpinEvolution} clearly demonstrates a linearly evolving $\dot\nu$, displaying a stable $\ddot\nu$. 
This stability is further illustrated in Fig. \ref{FigBreakingIndex}, which shows the instantaneous value $n(t)$ of the braking index as function of time.
These measurements show the resulting braking indices of fits of Eq. \ref{PowerLawTimingModel} to short stretches of data.
The fitting process is similar to that employed to produce Fig. \ref{FigSpinEvolution}, except that each data subset included all TOAs separated by less than half a year
from the central TOA. Stretches of data containing less than six TOAs
were excluded from the analysis. 
The derived $n(t)$ after Eq. \ref{eq:brakingindex}, from fits of Eq. \ref{EqBasicTimingModel} to the same stretches of data (not shown), are virtually identical. 
We stress here that the quoted statistical error on the braking index in Table \ref{TableTimingSolution} does not take fully into account the much larger systematic errors caused by timing noise (see \citealt{lnk+11}) which, as seen from the scatter of the instantaneous measurements in Fig. \ref{FigBreakingIndex}, is significant.
However, though clearly affected by timing noise, the measurements of the instantaneous value of $n(t)$ vary around a relatively well defined average. There is no evidence for a longer-term systematic evolution of the braking index before the first glitch. 

The glitches disrupt the gradual spin-down. The effect of the change in $\ddot{\nu}$ after the 2007 glitch on the measured instantaneous braking index $n(t)$, a decrease below its pre-glitch value as predicted by both model A and B, is clear in Figure \ref{FigBreakingIndex}. The measurements from fits in smaller subsets also indicate a similar decrease, despite their scatter due to timing noise.
According to model A this apparent decrease in $n(t)$ is transient and the instantaneous braking index is currently slowly recovering to its pre-glitch value. 
The permanent changes in model B however (see Table \ref{TableTimingSolution}), predict that the braking index will settle at a lower value, corresponding to a decrease in $n$ of about $\sim15\%$ after the 2007 glitch.
We stress that it is unclear if the glitch resulted in a permanent decrease of $n$. 
For example model A, which describes the data equally well, predicts that $n$ will recover to its original long-term value.

We cannot know for certain in what way the braking index is evolving in the long-term, however it is clear that the 2007 glitch resulted in a (possibly permanent) decrease in $n$.
As discussed further in section \ref{discussion} the timescales involved in the response of the superfluid stellar component can be very long. 
It is therefore rather likely that, at least in some pulsars, spin-down equilibrium is never reached between glitches.
If that is the case then $I_{\rm{ef{\kern0pt}f}}$ and the observed spin-down rate vary in a way that is not captured by the generalised Eq. \ref{obs:powerlaw}.
A time-dependent observed instantaneous braking index (as for example predicted by model A) is a strong indication of such an incomplete glitch recovery. 
The evolution prior to the observed glitches (to which the previously reported measurements of $n$ and the one quoted in Table \ref{TableTimingSolution} correspond) could also be ``contaminated" by the recovery of a glitch that occurred before our first observations. 
If such a glitch resulted also in a long-lasting decrease of the apparent braking index, we cannot rule out the possibility that the pre-glitch value of $n$ in Table \ref{TableTimingSolution} is underestimated. 
Potentially, the underlying braking index could even be consistent with $n=3$, the canonical value expected for the vacuum dipole braking mechanism.


\subsection{Limits on pulse profile shape variations}
\label{pulsechanges}

The pulse profile of PSR~J1119$-$6127 is usually single
peaked. Remarkably, the first observation after the 2007 glitch shows a double-peaked pulse profile. 
A double-peaked profile was not observed in any other observations analysed by WJE11 or in the subsequent data analysed in
this paper, making this event extremely rare. With the additional 77 new
observations analysed (in total 5.1 hours of data) the double-peaked
profile is observed in only 0.09\% of the total amount of data.  The
fact that this event is so rare suggests that it is not a coincidence
that the first observation after the 2007 glitch showed a
double-peaked profile, but rather that the two events are linked.

We also looked for more subtle shape variations than the above
discussed profile change. This was done by measuring the profile width
at 10\% ($W_{10}$) and 50\% ($W_{50}$) of the maximum intensity, by
fitting two von Mises functions to the individual observed profiles. 
No evidence of significant variations was found.
Moreover, to quantify any lower level profile shape changes throughout the long-timescale
recovery after the 2007 glitch, we summed the first and second half of
the post-glitch data separately before measuring their widths. 
No significant evolution could be identified within
a precision of 4\% in $W_{10}$ nor $W_{50}$. 
 Measuring the width from the summed data before and after the 2007 glitch (excluding
the observation with the double-peaked profile) did not reveal any significant profile
shape difference either\footnote{Formulating a formal uncertainty in this case is
  complicated, since the epoch of the glitch is close to the change
  from an analogue to a digital filter bank backend and a slight change
  in centre frequency. }.

It is known that timing noise, in particular $\dot{\nu}$ changes, can be related to profile shape changes
\citep{lhk+10}. We therefore tested for the presence of a correlation
between the variations seen in Fig. \ref{FigNuResiduals} and both
$W_{10}$ and $W_{50}$, but no such correlation was found. 

In addition to the change in the radio profile after the 2007 glitch, WJE11
report the appearance of abnormal erratic emission shortly after the 
glitch at
rotational phases where normally no radio emission is observed.
Since WJE11, only two 3 minute long observations (at a wavelength of 20 cm)
were recorded for which the individual pulses were stored. These 
observations (from April
and May 2012) do not show any evidence for similar erratic radio 
emission, thereby strengthening the claim that the previously
identified erratic behaviour is linked to the 2007 glitch.

\section{Physical interpretation of the post-glitch spin-down evolution}
\label{discussion}

Let us now explore the various physical mechanisms possibly involved in the ``over-recovery" of the spin-down rate after the 2007 glitch and its subsequent evolution. 
Ordinary glitches are attributed to the loose interaction of the NS's normal matter with the frictionless and irrotational neutron superfluid.
The rotation of the latter is supported by neutron vortices, which have quantised circulation $\kappa=h/2m_n$, where $h$ is the Planck constant and $2m_n$ is the mass of a neutron pair. 
Vortices carry the superfluid's angular momentum $L_s$ and are expected to be very dense in pulsars since their number density $n_{\rm{v}}$ is proportional to the superfluid's spin frequency.  
When free and in equilibrium, vortices arrange themselves in an array, mimicking solid body circulation for the superfluid.
The superfluid follows the rotation of its container by creation/expulsion of vortices and adjustments of their density. 

In the NS's interior the superfluid is immersed in normal matter and possibly coincides with a proton superconducting condensate in parts or the entire core. Thus the required adjustments of $n_{\rm{v}}$ might not always be possible, allowing differential rotation of the superfluid which alters the system's dynamics. 
The braking is then described by
\beq
\label{Ldot}
\frac{dL}{dt}=\int \dot{\Omega}_s(\vb{r},t)\,dI_s + I_c\,\dot{\Omega}_c
\eeq
which can be re-written as: 
\beq
\label{Next}
N_{\rm{EXT}}=\left(\frac{\int \dot{\Omega}_s(\vb{r},t)\,dI_s}{\dot{\Omega}_c} + I_c\right)\,\dot{\Omega}_c=I_{\rm{ef{\kern0pt}f}}\,\dot{\Omega}_c \; \text{.}
\eeq 
 Here $N_{\rm{EXT}}$ is the sum of all external torques, $L$ the angular momentum, $I$ denotes moment of inertia, $\Omega$ the angular velocity, indices `$s$' and `$c$' refer to the superfluid and the crust respectively and we have defined a new effective moment of inertia $I_{\rm{ef{\kern0pt}f}}$. 
The component $I_c$ includes the charged particles in the core and the magnetosphere, which are magnetically locked to the crust, and rotates at the pulsar's frequency $\nu=\Omega_c/2\pi$. 

The crystalline matter of the inner crust provides an inhomogeneous interacting potential for the vortices, as they are attracted -or repulsed- by the lattice nuclei. 
In the NS core, protons might form a type-II superconductor, in which case neutron vortices will interact with the lattice of proton flux-tubes.
Consequently NS vortices might not be free to move outwards in order to track the decreasing spin frequency of the star. 
If the interaction is strong vortices will be completely immobilised, ``pinned", and $\dot{\Omega}_s$ will be zero in this region. 
Smaller interaction energy $E_p$ will result in only partial restriction of the vortex flow, $|\dot{\Omega}_s|<|\dot{\Omega}_c|$. 
In both situations an excess of vorticity builds up locally, which translates as a rotational lag between the two components, $\omega(\vb{r},t)=\Omega_s(\vb{r},t)-\Omega_c(t)$.
In equilibrium, $\omega(\vb{r},t)=\omega_{\rm{eq}}(\vb{r},t)$ and the two fluids spin-down together ($\dot{\Omega}_c=\dot{\Omega}_s$).

The dif{\kern0pt}ferential rotation between a pinned vortex and the superfluid induces a lift (Magnus) force $F_m$ which counteracts pinning. In the axisymmetric case, $F_m\propto\rho_s r \omega$, where $\rho_s$ is the superfluid density and $r$ the cylindrical radius.  
If vortex outflow is restricted, the increasing lag $\omega$ will eventually reach a critical value $\omega_{\rm{cr}}$, defined by the maximum $F_m$ that can be sustained by the pinning force, unfreezing suddenly all excess vorticity. 
Such vortex avalanches might also be triggered locally by perturbations caused for example by fluid instabilities or a crustquake, and could lead to glitches of various sizes \citep{wm13}.  

In regions where pinning is not very strong, the thermal energy of vortices allows them to slowly escape outwards, hopping from one pinning site to another, which results in a temperature-dependent $\dot{\Omega}_s<0$. 
This idea of vortex ``creep" in NSs was firstly put forward by \citet{aaps84a}, who argued a temperature dependence of the form 
\[\dot{\Omega}_s\propto \exp\left[-\frac{E_p}{kT}\left(1-\frac{\omega}{\omega_{\rm{cr}}}\right)\right]-\exp\left[-\frac{E_p}{kT}\left(1+\frac{\omega}{\omega_{\rm{cr}}}\right)\right] \]
 for the creep rate in the case of thermally activated unpinning. 
Therefore the contribution of these regions to $I_{\rm{ef{\kern0pt}f}}$ can be highly sensitive to temperature changes. 
For very cool NSs, vortex unpinning by quantum tunnelling dominates the creep rate, which becomes almost independent of temperature \citep{bel92}. 

Vortices interact with both the superfluid and the normal component, via the Magnus force and a drag force respectively, exchanging angular momentum between the two. 
As in terrestrial superfluids, this coupling of the two components can be described by the mutual friction (MF) \citep{MF06}, which incorporates the effects of the vorticity and has a force density:
\beq 
 \label{MF1}
 \vb{f}=\rho_s \mathcal{B}'\bm{w}_s\times(\vb{v}_s-\vb{v}_c)+\rho_s\mathcal{B}\bm{\hat{w}}_s\times(\bm{w}_s \times(\vb{v}_s-\vb{v}_c))   \;\text{.}
 \eeq
We have introduced the superfluid vorticity $\bm{w}_s=n_{\rm{v}}\kappa\bm{\hat{w}_s}$, where $\bm{\hat{w}}_s$ is a unit vector aligned with the vortex axis, and the MF coefficients $\mathcal{B}'$ and $\mathcal{B}$.
 
While all vortices, either pinned or free, contribute to the superfluid's velocity $\vb{v}_s$, only unpinned vortices move with respect to the normal fluid and hence feel the drag force. 
Therefore the above coefficients $\mathcal{B}'$ and $\mathcal{B}$ depend not only on the strength of the coupling mechanism 
 but also on the fraction $\xi$ of unpinned vortices, which is a function of the temperature $T$ and the lag $\omega$ relatively to $\omega_{\rm{cr}}$. 
This fraction will be proportional to the probability for unpinning from an energy barrier of $\Delta E=E_p(\omega_{\rm{cr}}-\omega)/\omega_{\rm{cr}}$, therefore for thermal unpinning $\xi\propto\exp(-\Delta E/kT)$.

If a significant increase in coupling strength happens quickly, due to catastrophic unpinning (vortex avalanche, $\xi\rightarrow 1$) and fast dissipation, the superfluid accelerates the crust causing a glitch.  
Immediately after the spin-up of the crust the lag decreases, leading to temporarily decoupling of the superfluid, a decrease in $I_{\rm{ef{\kern0pt}f}}$ and the observed post-glitch recovery. 

Permanent changes in spin and spin-down rate that are sometimes seen after glitches \citep{Yu13} have been attributed to magnetic axis re-orientation or/and moment of inertia changes, due to NS crustquakes \citep{linkcrab92,Acrab94,Rud98}.
Crustquakes are expected if the solid crust does not plastically adjust to the less oblate equilibrium shape required by the pulsar's spin-down. 
In this case the most natural outcome of the readjustment is a decrease in moment of inertia\footnote{This however might not be the case for a magnetically strained crust.} $\Delta I<0$, which results in permanent changes in rotation $\Delta\nu_p/\nu=\Delta\dot{\nu}_p/\dot{\nu}=-\Delta I/I$.
Such abrupt events could trigger the unpinning of vortices and have been connected to both glitches and outbursts (e.g., \cite{LEp96,PonsRea12}). 

We will adopt the above picture of an abrupt unpinning event (either supplemented by a crustquake episode or not) for the origin of both PSR~J1119$-$6127 glitches.
This is motivated by their striking similarity, in both the initial glitch parameters, $\Delta\nu$ and $\Delta\dot{\nu}$, as well as the first part of the relaxation process, to regular radio pulsar glitches. 
However this mechanism alone cannot explain the ``overshooting" of the pre-glitch $\dot{\nu}$ ($\Delta\dot{\nu}>0$). 
Moreover, it might result in persistent $\Delta\dot{\nu}<0$ because of the long recoupling timescales of the crustal superfluid \citep{link14,bryn14}.
Consequently the subsequent $\dot{\nu}$ evolution must be governed by a different process. 

The regular glitch relaxation (with $\Delta\dot{\nu}<0$) masks the behaviour of the ``irregular" term for the first few years post-glitch, therefore it is not possible to obtain a clear description for this period.
Nonetheless, the maximum observed positive $\dot{\nu}$ deviation $\Delta\dot{\nu}_{\rm{max}}\simeq3.5\times10^{-14}\,\rm{Hz/s}$ provides a lower limit for the amplitude of the decrease in spin-down rate, which we use to examine the plausibility of various physical mechanisms for the post-glitch variation.  

We first discuss internal (superfluid) mechanisms that could be responsible for the strange long-term recovery of $\dot{\nu}$, in section \ref{superfluid}.  
The peculiar radio emission features observed after the 2007 glitch recovered on a rather fast timescale ($\lesssim 3$ months) compared to the timescales characterising the long-term $\dot{\nu}$ evolution.
Nevertheless, it indicates magnetospheric activity related to the glitch so changes in $N_{\rm{EXT}}$ cannot be excluded.
This possibility is further explored in section \ref{magnetosphere}.   
As we will show, small changes in magnetospheric conditions suffice to explain the observed variation in $|\dot{\nu}|$. 
The consequences  of such changes in the radio emission could have been small enough to be missed entirely, if it were not for the one observation of the double-peaked profile right after the glitch.  



\subsection{Internal, superfluid mechanisms} 
\label{superfluid}
\indent

We can think of two ways in which the superfluid might be responsible for the observed $\dot{\nu}$ evolution. 
First, it could be that the spin-up glitch was accompanied by an ``anti-glitch", that is, a fast spin-down event.  
In this scenario some vortices move inwards rather than outwards during the glitch, causing a spin-up of the respective superfluid region $I_{\rm{in}}$ and an increase $\Delta\omega(I_{\rm{in}})=\Delta\omega_{\rm{in}}>0$ of the local lag. 
For simplicity we will consider the case where this region does not coincide with the superfluid region that drives the glitch. 
The lag of the region $I_{\rm{in}}$ would decrease by $\Delta\omega_{\rm{g}}\sim-2\pi\Delta\nu_{\rm{g}}$ because of the glitch, in the absence of a vortex flow. 
The net effect on the lag, and therefore the coupling strength of this region with the normal component and its contribution to $I_{\rm{ef{\kern0pt}f}}$, will depend on the ratio $|\Delta\omega_{\rm{in}}/\Delta\omega_{\rm{g}}|$. 
Thus such a picture can lead to a large variety of features in the timing residuals, depending on the above ratio and the size and local coupling timescale of the region where vortex inflow took place.  

However an inwards vortex motion is not favoured for a spinning-down NS, thus an additional mechanism to the ones discussed above must be invoked.  
\citet{accp96} proposed the formation of a new pinning region (vortex trap) as a possible explanation for a similar, peculiar glitch feature observed in the Crab pulsar \citep{lsp92}, where a gradual increase in $\Delta\nu$ (relative decrease of $|\dot{\nu}|$) followed a normal (sudden) glitch. 
If in the creep regime the inward vortex motion is not largely suppressed, some vortices will end up pinned in this ``trap" at a distance closer to the rotational axis than they were before.  
The formation of a new ``trap" would leave a permanent imprint in the spin-down rate $\Delta\dot{\nu}_p<0$, since $I_{\rm{ef{\kern0pt}f}}$ reduces. 
This is indeed observed in the Crab, but is not in accordance to the $\dot{\nu}$ evolution seen so far in PSR~J1119$-$6127.

It could be that vortices were instead forced to move inwards because of their interaction with the crustal lattice or the core's flux-tubes, to a region where pinning properties remained unchanged. 
In this case, if $\Delta\omega_{\rm{in}}/\Delta\omega_{\rm{g}}>1$ then this region contributes {\it more} to $I_{\rm{ef{\kern0pt}f}}$ after the glitch (reflected in the post-glitch relaxation as a relaxing component with $\Delta\dot{\nu}>0$, as in model A).
According to the parameters derived from the data (see Table \ref{TableTimingSolution}), the region where this happens must have a very long coupling timescale, of the order of years as indicated by $\tau_d^{(3)}$.
This scenario is examined in detail by Akbal et al. (submitted to MNRAS), who extend the phenomenological creep model of \citet{aaps84a} to account for a region with vortex inflow at the time of the glitch.
Because of the non-linear creep term included in their 9-parameter fits, the derived parameters do not correspond directly to the ones of model A. However, a permanent change in the external torque of similar magnitude to our calculations (see sections \ref{observations} and \ref{magnetosphere}) is also required by their model. 

There is a different way in which the superfluid could be responsible for the post-glitch $\dot{\nu}$ evolution, that does not require inwards vortex motion.
Rather, the temporal enhancement of the coupling strength between parts of the superfluid and the normal component can be attributed to local temperature variations following the glitch, instead of an increase of their lag.
From Eq. \ref{MF1} and in the axisymmetric case, the superfluid spin-down rate $\dot{\Omega}_s(\vb{r},t)$ in Eqs. \ref{Ldot} and \ref{Next} can be written as \citep{trevor10}
\beq
\label{nusdotMF}
 \dot{\Omega}_s(\vb{r},t)= \mathcal{B}(\vb{r})|w_s(\vb{r},t)|(\Omega_c(t)-\Omega_s(\vb{r},t))
 \eeq
where for simplicity we ignore the effects of entrainment. 
Increased vortex mobility, expressed as a higher value $\xi$ and therefore larger $\mathcal{B}$, leads to stronger coupling between the superfluid and normal component for a given lag $\omega$ and to a temporal increase of $I_{\rm{ef{\kern0pt}f}}$ (Eq. \ref{Next}).
This mechanism offers a promising possibility for explaining the observed long lasting $\dot{\nu}$ evolution, since both the NS cooling and the superfluid response are relatively slow processes. 
Furthermore, such changes in coupling strength have been shown to lead to observable effects on $\dot{\nu}$ very similar to the irregular feature of the PSR~J1119$-$6127 glitches \citep{bryn14}.

During the glitch, energy is dissipated due to vortex motion, while crustquakes and radiative bursts can release elastic and magnetic energy stored in the strained crust and magnetic field lines. 
Therefore some heat input in the inner crust coincidental with the glitch is to be expected, which can be as large as $10^{40}-10^{44} \rm{erg}$ if additional processes such as a crustquake are involved \citep{pernapons11}.
Part of this energy is transferred to the magnetosphere on fast timescales, and it could be responsible for the activation of the radio emission mechanism in a region of previously inactive magnetic field lines, producing the observed changes in the pulse profile. 
Depending on the temperature and physical properties of the region where the energy was injected, part of it will be lost quickly via neutrino emission and in the highly conductive core, while the rest will cause local heating, eventually being radiated as thermal emission from the surface on much longer timescales \citep[see for example][]{Aguilera08}. 
A strong toroidal magnetic field around the crust-core boundary, as inferred for PSR~J1119$-$6127 \citep{ng12}, might help the confinement of heat in this region.  

The energy release can be regarded as ``instantaneous" since the injection timescale for most plausible mechanisms (glitch dissipation, crustquake or magnetic reconnection) is much smaller than the timescales of interest (for the $\dot{\nu}$ behaviour). 
In the vicinity of the energy source the temperature will start growing and the extent of the heated region $I_{\rm{H}}(t)$ will increase with time. 
The ef{\kern0pt}fect of the reduced lag $\omega_{\rm{H}}$ of this region because of the glitch might be stronger than the change in coupling due to a slightly higher temperature. Thus initially the net ef{\kern0pt}fect can be a negative change in $\dot{\nu}$.
As the superfluid re-couples though, the relaxation will be towards the new pseudo-equilibrium lag $\omega_{\rm{eq}, \rm{H}}(t)$ of the heated area, which will remain smaller than the pre-glitch one during the cooling back to the pre-glitch temperature profile. 
As a consequence, the contribution to the ef{\kern0pt}fective moment of inertia $I_{\rm{ef{\kern0pt}f}}$ of this region will be larger after the glitch than it would have been had the temperature remained constant. This is reflected in the positive $\Delta\dot{\nu}$. 
The observed deviation $\Delta\dot{\nu}_{\rm{max}}$ provides a lower limit for the extra contribution of the heated region in $I_{\rm{ef{\kern0pt}f}}$.
The equivalent change in $\mathcal{B}$ depends on the size $I_{\rm{H}}(t)$ of the af{\kern0pt}fected region, as well as on its properties (as depth, pinning/drag strength and initial temperature) which define $\omega_{\rm{eq}}$. 
Note that for the above mechanism to work, the temperature increase must happen over a region that was partially decoupled ($\dot{\Omega}_s\neq\dot{\Omega}_c$) before the glitch. 

In this scenario the glitch will not be accompanied by a sudden spin-down (as parameterised in model A), unlike the previously discussed mechanism.
Instead, $I_{\rm{ef{\kern0pt}f}}$ (and thus $\Delta\dot{\nu}$) will gradually reach a maximum followed by a decrease at a rate defined by the post-glitch cooling rate. 
This decrease might be reflected in the $\Delta\ddot{\nu}_p$ term of model B and the corresponding change in the braking index. 
Near the end of the cooling phase however, a turn in $\Delta\dot{\nu}$ is expected, which will either fully recover to the projected pre-glitch $\dot{\nu}$ ($\Delta\dot{\nu}=0$) or, in the presence of a change in the external torque, to a new stable state.  
The functional form of the relaxation is defined by the global response of the superfluid to the glitch, the temperature evolution $T(t)$ as well as the exact form of the dependence of the dissipative mutual friction coefficient $\mathcal{B}$ on $T$, which is unknown. 
If such a mechanism is at work, glitches where this signature is clear in the timing residuals can be very useful probes of $\mathcal{B}(T)$. 
Since the neutron superfluid forms very soon after the NS birth, the dependence of $\dot{\Omega}_s$ on $T$ could also have an impact on the early rotational dynamics of pulsars, during the fast cooling phase.



\subsection{Magnetospheric mechanisms}
\label{magnetosphere}

We will now focus on external, magnetospheric mechanisms that could lead to the peculiar spin-down rate evolution. 
Such mechanisms have observational corollaries and can potentially be tested by future high-energy observations.

In the simplest approximation for pulsar spin-down, the vacuum dipole model, the magnetospheric torque on the star is given by   
\beq
\label{vactorque}
N_{\rm{vd}}=-\frac{{R_{\star}}^6\Omega^3}{6c^3}B_{\star}^2\sin^2{\alpha}
\eeq
 where $B_{\star}$ is the surface magnetic field at the pole, $R_{\star}$ the stellar radius and $\alpha$ the inclination angle. 
 
 Pulsars are not surrounded by vacuum as rotation induces an electric field $ \vb{E}=-(\bm{\Omega}_c\times\vb{r})\times\vb{B} $, which accelerates charges off the neutron star surface. 
Secondary pair creation from those charges and cascades are expected to fill the magnetosphere with plasma, which will screen the electric field along the magnetic field lines (hereafter denoted $E_{\parallel}$).
The currents that flow in the magnetosphere and close under the star's surface provide an additional braking torque which is comparable to the one of Eq. \ref{vactorque} \citep{harding99} and therefore should not be neglected. 

The plasma density required to completely screen $E_{\parallel}$ is the Goldreich-Julian density, $\rho_{\rm{GJ}}=-B_{\star} \nu/c$ \citep{GJ69}.
If plasma is abundant, the magnetosphere can be considered in the force-free regime, where $E_{\parallel}=0$ everywhere except in the acceleration zones above the polar caps and regions where plasma flow is required, like the equatorial current sheet.  
The drift velocity of the magnetospheric plasma has a rotational component $\Omega_F$, which represents the angular velocity of the magnetic field lines. 
For an aligned rotator, which is the approximation we will mostly use for numerical calculations in the following, $\bm{\Omega}_F=\Omega_F \bm{\hat{e}}_z$. 
The behaviour of the magnetosphere will be qualitatively the same for inclination angles as the ones inferred for PSR~J1119$-$6127 ($\alpha\sim17^{\circ}-30^{\circ}$, WJE11). 
In the open field line region $\Omega_F<\Omega_c$ and is determined by the potential drop along the magnetic field lines, which depends on the poorly understood microphysics of the cascade zone. 
Different $\Omega_F$ and poloidal current density distributions lead to different global magnetospheric structure and Poynting flux to infinity.
The energy losses, and thus the additional braking $\dot{\nu}$, depend in general on $\Omega_F\Omega_c$ and the size of the open field line region \citep{Cont05,ContSpit06,Timokhin06}. 

The spin-down rate for a force-free magnetosphere was shown to be $\gtrsim 3$ times larger than that for a misaligned rotator in vacuum. 
In this regime simulations by \citet{Spit06} show a dependence of the torque on the inclination angle of the form: 
 \beq
\label{spit06}
N_{\rm{f\,f}}\simeq-\frac{{B^2 R_{\star}}^6\Omega^3}{c^3}(1+\sin^2{\alpha}) \; \text{ .}
\eeq

If the relative decrease in $|\dot{\nu}|$ of PSR~J1119$-$6127 was due to a glitch-induced change in the surface magnetic field, then it would imply, for a rotator in vacuum, a decrease in the dipole component $B_d$ of the order $\Delta B_d\sim3\times10^{10}\rm{G}$.
An actual decrease of the surface magnetic field strength would have released energy $\sim\Delta B\times B/8\pi\simeq5\times10^{22} \; \rm{erg}$, however a more likely explanation for a decrease of $B_d$ is a change in the inclination angle $\alpha$. 
Such a geometric readjustment is possible if the glitch was accompanied by crust failure.
In order for the pulsar to become more spherical at a crustquake platelets should move towards the rotational axis (see section \ref{superfluid}).
The magnetic field lines, anchored at the highly conductive crust, follow this motion, thus $\alpha$ can decrease. 
The frequency jump at the glitch places an upper limit in the respective decrease in moment of inertia ${\Delta I}< I_c \Delta\nu/\nu$ and has a small impact on the spin-down rate $(\Delta\dot{\nu}/\dot{\nu})_{\rm{quake}}\lesssim3\times10^{-5}$. 
Therefore the overall ef{\kern0pt}fect can be a decrease in spin-down rate, as observed.

Such structural changes will result in a permanent $\dot{\nu}$ change instead of a decaying $\Delta\dot{\nu}>0$ (as in model A). 
Nevertheless, they cannot be ruled out and could potentially explain the permanent $\Delta\dot{\nu}_p$ required by model B, or the smaller residual $\Delta\dot{\nu}\sim23\%\Delta\dot{\nu}_p$ to which the solution of model A tends asymptotically.  
For a small change $\Delta\alpha$ of the inclination angle, the relative change in torque is \(\,\Delta N_{\rm{vd}}/N_{\rm{vd}}\simeq2\Delta \alpha/\tan{\alpha}\,\) for the vacuum dipole model, or according to Eq. \ref{spit06} 
\[\Delta N_{\rm{f\,f}}/N_{\rm{f\,f}}\simeq\Delta \alpha \sin{2\alpha}/(1+\sin^2{\alpha}) \; \text{.}\] 
From model B, $\Delta N_{\rm{EXT}}/N_{\rm{EXT}}=\Delta\dot{\nu}_p/\dot{\nu}=-13.5\times10^{-4}$. 
Using the inferred possible range of inclination angles for PSR~J1119$-$6127, the equivalent decrease in inclination angle is found $\Delta \alpha\sim-(3.9-2)\times10^{-4}$ or $\Delta\alpha\sim-(2.6-1.9)\times10^{-3}$ degrees according to Eqs. \ref{vactorque} and \ref{spit06} respectively. 
The ef{\kern0pt}fects of such a change of $\alpha$ on the pulse profile cannot be excluded by the observations (see section \ref{pulsechanges}).   

The appearance of the extremely rare double peaked radio profile in PSR~J1119$-$6127 indicates some perturbation of the pre-glitch magnetospheric state, as a direct or indirect consequence of the glitch.  
The first post-glitch observations suggest global magnetospheric changes which reflect on the radio emission (WJE11). 
There is growing evidence for similar changes, as well as a correlation between radio emission characteristics and spin-down rate, in magnetospheric active pulsars like mode changing and intermittent pulsars \citep{kramer06Sci,lhk+10,crcjd12,J1832Lorimer12,B1931Young13}. 
Thus it appears that while some pulsars have relatively stable magnetospheric states, others can switch between metastable states, possibly within a few spin periods \citep{kramer06Sci}. 
This magnetospheric state switching could be induced by a glitch, as suggested not only by PSR~J1119$-$6127, but also by the observations of the mode changing PSR~J0742$-$2822. 
For this pulsar, \citet{Keith13} reported a stronger correlation between radio emission mode and $\dot{\nu}$ after a glitch.

The observed $|\Delta\dot{\nu}|_{\rm{max}}$ of the PSR~J1119$-$6127 glitch recovery requires small changes (compared to those observed in intermittent pulsars for example) around its pre-glitch magnetospheric state, which appears stable overall. 
It could be that PSR~J1119$-$6127 is delicately balanced in this state and the glitch triggered a short-lived transition to a different one, or the permanent transition to another quasi-steady state of similar characteristics. 
The change of the magnetospheric torque when the profile was double peaked could have been more dramatic, but $\dot{\nu}$ was still dominated by the glitch relaxation at that time.  
A similar mechanism might be at work in other glitching pulsars too but, due to (1) smaller glitch sizes, (2) geometric effects or (3) more stable magnetospheric states and faster relaxation timescales, we would not necessarily see prominent  pulse profile changes \citep{tim10} and the magnetosphere-related decreases in spin-down rate might be masked by the post-glitch relaxations. 

The glitch (and/or a simultaneous crustquake) could have triggered a plasma deficiency in the magnetosphere, for example by altering temporarily the properties in the cascade zone.  
As coherent radio emission is expected to be sensitive to those properties, such a change might explain the short-lived erratic components and the double-peaked profile seen right after the glitch, as well as the transient decrease in spin-down rate. 
After the glitch, the plasma supply is restored and the magnetosphere recovers to another stable state, on a short (model B) or long timescale (model A). 
In the first case (model B), the magnetospheric changes are reflected in the permanent terms $\Delta\dot{\nu}_p$ and $\Delta\ddot{\nu}_p$.  
Even though intuitively the transition is expected to be achieved in the rather fast timescales that characterise the magnetosphere (of the order of $\mu\rm{sec}$ for the cascade zone and up to $\sim 1$ year for Ohmic decay \citep{rudsuth75,BT07corona}), non-linear processes involving global magnetospheric adjustment (from the polar cap to the rest of the magnetosphere and the current sheet) can lead to much longer timescales. 
Thus model A cannot be excluded, as recovery to a new stable state might be slow.  
Alternatively, possible glitch-induced reconnection of magnetic field lines at distances $\gtrsim R_{\rm{LC}}$, where $R_{\rm{LC}}$ is the light cylinder radius\footnote{Since $R_{\rm{LC}}\sim c/(2\pi\nu)\sim 2\times 10^{3} R_\star$, the magnetic field there is considerably weaker than at the surface.}, which alters the return current and the magnetospheric structure, might lead to similar results. 

A simple estimate for the decrease in $|\dot{\nu}|$ due to plasma shortage can be obtained if we assume that the positive $\Delta\dot{\nu}_{\rm{max}}$ is achieved when the charge density $\rho=0$ (vacuum regime, Eq. \ref{vactorque}), while in the pre-glitch state there is an additional braking torque, proportional to the current through the polar cap. 
Then the plasma density in the pre-glitch state should have been of the order  \citep{harding99}
\beq
\label{plasmadensity}
\rho=\frac{3I\Delta\dot{\nu}}{R_{\rm{PC}}^4 B_\star}=\frac{3c^2 I \Delta\dot{\nu}}{4 \pi^2 R_{\star}^6B_{\star}\nu^2}\sim10 \;\rm{statC\,cm^{-3}}
\eeq
where we used $R_{\rm{PC}}\simeq (2\pi R_\star^3 \nu/c)^{1/2}$ as an approximation for the polar cap radius, $\Delta\dot{\nu}=\Delta\dot{\nu}_{\rm{max}}$, $R_{\star}\sim10^{6}\,\rm{cm}$ and $I\sim10^{45}\;\rm{g\,cm^2}$. 
This is a very small fraction of $\rho_{GJ}$ (less than $1\%$). 
It is more natural to assume that the seemingly stable pre-glitch state is closer to the force-free regime (which seems energetically favourable and will have a plasma density comparable to $\rho_{GJ}$) and that the magnetosphere does not deplete completely from plasma in the low $|\dot{\nu}|$ state.  
In this case the observed evolution of $\dot{\nu}$ reflects the build-up of plasma density, the effect of which on spin-down is investigated (indirectly, by varying the conductivity of the magnetosphere) in \citet{Li12}.

Another way to understand this is that when charges are not copiously available, the magnetosphere is unable to support a poloidal electric current as large as before. 
Therefore a state with smaller open field regions and different current and toroidal $B_{\phi}$ configurations is energetically favourable \citep{Cont05,ContSpit06}.
In this state the Poynting flux to infinity is smaller, and so is the magnetospheric torque and $|\dot{\nu}|$. 

In the case of PSR~J1119$-$6127, this new magnetospheric state might have been achieved for example via reconnection of open field lines close to the light cylinder quickly after the glitch, and would appear as a positive $\Delta\dot{\nu}$. 
Once the plasma supply is restored and the poloidal current increases again, the magnetosphere will evolve to a state with larger open field line region, which does not necessarily coincide with the pre-glitch state. 
The timescale of this recovery will depend amongst other things on the rate in which energy is stored in the magnetosphere (and will thus vary from pulsar to pulsar).

For an aligned rotator, \citet{Cont05} investigated the above effects of charge deficiency on $\Omega_F$, and the subsequent decrease of electric currents, on the magnetospheric structure and spin-down rate.
The difference in magnetospheric energy contained within a cylindrical radius $r$, for two different $\Omega_F$ of the open field line regions $\Omega_{F,1}$ and $\Omega_{F,2}$, is (see Eq. 24 in \citet{Cont05})
 \beq
 \Delta E_{\rm{EM}}\sim B^2 R_{\star}^3\left(\frac{R_{\star}}{R_{\rm{LC}}}\right)^3\left(\frac{r}{R_{\rm{LC}}}\right)(\Omega_{F,2}^2-\Omega_{F,1}^2) \; \text{ .}
 \eeq
 The ratio of the respective spin down rates $\dot{\nu}_1$ and $\dot{\nu}_2$ of those states (with $\Omega_1\simeq\Omega_2=\Omega_c$) is
 \beq
 \label{cont05}
 \frac{\dot{\nu}_1}{\dot{\nu_2}}=\frac{\Omega_1^2\Omega_{F,1}}{\Omega_2^2\Omega_{F,2}}\simeq\frac{\Omega_{F,1}}{\Omega_{F,2}} \;\text{ . }
 \eeq
 We can make a crude approximation by assuming an initial state with $\Omega_{F,1}=0.803\,\Omega_c$ and using $\dot{\nu}_2=\dot{\nu}_1+\Delta\dot{\nu}_{\rm{max}}$. This leads to $\Omega_{F,2}=0.802\,\Omega_c$ for a NS with mass $1M_\odot$ and $R_\star=10^6 \rm{cm}$. The released energy is then  
 \beq
 \Delta E\sim 10^{44} \left(\frac{R_{\star}}{R_{\rm{LC}}}\right)^3\left(\frac{r}{R_{\rm{LC}}}\right) \,\rm{ergs}\,\text{,}
 \eeq
thus for $r\simeq R_{\rm{LC}}$, $\Delta E \sim10^{34} \,\rm{ergs}$. 
 Note however that this quantitative approach is just indicative, since, according to the spin-down luminosity calculated by \citet{Cont05}, the condition $0<\Omega_F\leq\Omega_c$ implies a large and rather unlikely underestimation of the real magnetic field of PSR~J1119$-$6127, which should be more than an order of magnitude higher than the characteristic magnetic field inferred from Eq. \ref{vactorque}.


\section{Discussion and conclusions}
\label{sumperasmata}

Continuous monitoring of PSR J1119$-$6127 for $7$ years after its 2007 glitch, which was followed by an abnormal post-glitch recovery and glitch-induced emission changes, reveals an ongoing evolution of the residual post-glitch $\Delta\dot{\nu}$. 
This allows us to exclude structural adjustments as the main cause of the peculiar post-glitch feature in spin-down rate, since such changes, like a change in inclination angle $\alpha$ or the weakening of  vortex ``pinning" in some crustal region due to lattice failure, are expected to result in a permanent shift in $\dot{\nu}$. 

While the thermal output of some internal mechanisms (section \ref{superfluid}) might be below detectable levels, the external processes discussed in section \ref{magnetosphere} could have a significant imprint on the emission. 
A positive $\Delta\dot{\nu}$ was observed in both the large and moderate glitch of this pulsar, so it might be a common feature of its post-glitch spin-down behaviour. 
Broadband regular monitoring, from radio to $\gamma$-rays, and pointed X-ray observations as soon as possible after its next glitch to follow any thermal evolution, could allow us to clarify whether internal or external (magnetospheric) mechanisms dominate its complicated post-glitch spin-down.  

Glitch associated radiative changes and rotational features like the ones discussed here are observed mostly in magnetars. Similar behaviour might be present in typical NSs although not always detectable. 
For example, in the case of the Crab pulsar the relaxation after some of its glitches seems to be better described  when a timing model with an exponentially decaying $\Delta\dot{\nu}>0$ term is included \citep{ljg+14}. 
Nevertheless, in the Crab pulsar this feature is weaker than in PSR J1119$-$6127 so that $|\dot{\nu}|$ stays always larger than the pre-glitch projected value. 

Some of these differences in glitch phenomenology are possibly related to the magnetic field strength and structure. 
The X-ray pulse profile of PSR J1119$-$6127 is indicative of a strong toroidal component in the interior \citep{ng12} which might be still evolving in the crust \citep{vigano13}, as is also suspected to be the case in magnetars. Analysis of its braking index, employing a self-consistent and phase coherent technique, indicates a long-lasting apparent decrease after the 2007 glitch. 
The extrapolated evolution in the $P-\dot{P}$ plane according to the pre-glitch braking index would bring  PSR J1119$-$6127 to a region populated by long-period RRATs and magnetars.
If the glitch resulted in a new, lower long-term braking index then the inferred magnetic field would appear to be growing at a faster rate than before the glitch, and the $P-\dot{P}$ track would turn more towards the magnetar region.   

A post-glitch evolution characterised by a decrease in braking index has been observed before in only one other rotationally powered pulsar, the X-ray pulsar PSR~J1846$-$0258, which might also be evolving towards the magnetar population. 
This pulsar shares some interesting properties with PSR~J1119$-$6127, as it is also a very young ($\sim0.8\,\rm{kyr}$) pulsar with a high inferred magnetic field of $\sim5\times10^{13}\,\rm{G}$.
 Moreover, like the second glitch of PSR~J1119$-$6127, the 2006 glitch of PSR~J1846$-$0258 was relatively large in magnitude for such a young pulsar, and accompanied by radiative changes, with a X-ray flux increase and several X-ray bursts, an episode similar to magnetar outbursts \citep{ggg+08}.
 Following this glitch, a decrease in the braking index of $18\%\pm5\%$ was observed \citep{lnk+11}.
However the phenomenology of the glitch recoveries for these two pulsars appears different, possibly due to differences in the characteristic timescales of the underlying physical mechanism(s) and the amplitudes of the glitch-induced changes. 
In the case of PSR~J1846$-$0258 the post-glitch evolution is dominated by a very large increase in $|\dot{\nu}|$, which results in an ``over-recovery" in frequency, parameterised as a large ($\sim-10^{-4}\,\rm{Hz}$) permanent decrease in $\nu$ \citep{lkg10}.
In both pulsars the part of the glitch recovery modelled as exponential relaxations on short and intermediate timescales is over after about two years.  
However, the post-glitch $|\dot{\nu}|$ of PSR~J1846$-$0258 remains larger than the pre-glitch projected value for at least four years \citep{lnk+11} and is dominated by the glitch-induced enhancement of the spin-down rate (as in the case of the Crab pulsar mentioned above).  
This persistent faster spin-down is rather common for glitches in general, but is in clear contrast with the remarkable ``over-recovery" of $\dot{\nu}$ observed in PSR~J1119$-$6127. 
Nonetheless, in both cases there appears to be a prolonged decrease in $\ddot{\nu}$, which becomes detectable once the short-term recovery is no longer dominant, and is the main factor responsible for the apparent decrease in the braking index. 

The RRAT PSR J1819$-$1458 is the only other radio rotationally-powered pulsar with a post-glitch $\dot{\nu}$ evolution that results in a relative decrease in $|\dot{\nu}|$, and it is also a high magnetic field pulsar, with inferred $B_d\simeq5\times10^{13}\,\rm{G}$.
The recovery after its second observed glitch demonstrates the same peculiar characteristics as in PSR J1119$-$6127, with a measured persistent change $\Delta\dot{\nu}_p\sim0.8\times10^{-15}\,\rm{Hz\,s^{-1}}$.  
Interestingly, there are also indications for augmented activity of the radio emission following the glitches in PSR J1819$-$1458 \citep{lmk+10}. 
However, RRATs are characterised by their irregular radio emission, while PSR J1119$-$6127  displays a very stable radio emission and pulse profile during all other observations in the 16 years of monitoring. 
With emission characteristics much closer to those of normal, steady radio pulsars and timing noise levels considerably lower than in magnetars, timing studies of PSR J1119$-$6127 and similar high magnetic field pulsars might offer the possibility of disentangling the contribution of the magnetosphere and the neutron superfluid in the glitch-bursting puzzle.

\section*{Acknowledgments}
We thank the anonymous referee for constructive comments and Onur Akbal for an advance copy of their manuscript. 
D.A. would also like to thank Ali Alpar and Daniele Vigan{\`o} for useful discussions.
D.A. and A.L.W.  acknowledge support from an NWO Vidi Grant (PI Watts).
Pulsar research at JBCA is supported by a Consolidated Grant from the UK Science and Technology Facilities Council (STFC). 
C.M.E. acknowledges support from FONDECYT (postdoctorado 3130512).
The Parkes radio telescope is part of the Australia Telescope which is funded by the Commonwealth of Australia for operation as a National Facility managed by CSIRO.

\bibliographystyle{mn2e-mod}
\bibliography{journals,modrefs,psrrefs,crossrefs,new1,refs}

\label{lastpage}
\end{document}